\documentstyle[seceq,preprint]{ptptex}

\preprintnumber[3cm]{
IFUP-TH-50-99\\
KOBE-TH-99-05\\
YITP-99-55\\
hep-th/9909076}

\title{
Many-Brane Extension of the Randall-Sundrum Solution
}

\author{
Hisaki {\sc HATANAKA}$^{1,}$
\footnote{e-mail : hatanaka@phys.sci.kobe-u.ac.jp},
Makoto {\sc SAKAMOTO}$^{2,}$
\footnote{e-mail : sakamoto@phys.sci.kobe-u.ac.jp},\\
Motoi {\sc TACHIBANA}$^{3,}$
\footnote{e-mail : motoi@yukawa.kyoto-u.ac.jp} 
and Kazunori {\sc TAKENAGA}$^{4,}$
\footnote{e-mail : takenaga@ibmth.df.unipi.it}
}

\inst{
$^1$Graduate School of Science and Technology, Kobe University,\\
Kobe 657-8501, Japan
\\
$^{2}$Department of Physics, Kobe University, Kobe 657-8501, Japan\\
$^{3}$Yukawa Institute for Theoretical Physics,\\
Kyoto University, Kyoto 606-8502, Japan \\
$^{4}$I.N.F.N., Sezione di Pisa,
Buonarotti 2, Ed. B, 56127 Pisa, Italy}

\recdate{
September 16, 1999
}

\abst{
Recently, Randall and Sundrum proposed a static solution to Einstein's
equations in five spacetime dimensions with two 3-branes located at the
fixed points of $S^1/Z_2$ to solve the hierarchy problem. We extend the
solution and construct static and also inflationary solutions to
Einstein's equations in five spacetime dimensions, one of which is
compactified on $S^1$, with any number of 3-branes whose locations are
taken to be arbitrary. We discuss how the hierarchy problem can be
explained in our model.
}

\begin{document}

\maketitle

\noindent {\it 1. Introduction}
\hspace{5mm}
Recently, Randall and Sundrum \cite{Randall:1999ee} proposed a new
interesting mechanism with a single small extra dimension for solving the
hierarchy problem between the Planck scale and the weak scale. A key
ingredient of this mechanism is that the metric is not factorizable and
that the 4-dimensional metric is multiplied by a warp factor which is
a rapidly changing function of the extra dimension. They explicitly
constructed such a solution to the Einstein equations in 5 spacetime
dimensions, one of which is compactified on $S^1/Z_2$, with two 3-branes
located at the fixed points of $S^1/Z_2$. For a solution to exist, it is
crucial to take into account the effect of the branes on the bulk
gravitational metric.

In the Randall-Sundrum model, the number of 3-branes is 2 and the
locations are taken to be the fixed points of $S^1/Z_2$. Although this
setup is motivated by recent developments in string and M-theory,
\cite{horava} it would be of interest to construct new solutions to
5-dimensional the Einstei equations with many 3-branes. In this
letter, we explicitly construct solutions with an arbitrary number 
of 3-branes which are put at arbitrary positions in the direction of the 
extra dimension.
Our motivation for this is threefold. First, it is known that any number
of parallel D-branes can be put at arbitrary positions and that the gauge
dynamics depend on the distances of multiple
D-branes. Thus, it is physically meaningful
to construct solutions corresponding to many 3-branes put at arbitrary
positions.
Second, many-brane configurations could explain other hierarchy problems,
such as the fermion mass hierarchy. The original Randall-Sundrum model
gives a solution to the hierarchy problem between the Planck scale and
the weak scale, but it does not solve other hierarchy problems.
In recent works on large extra dimensions,\cite{LED} various mechanisms
to solve hierarchy problems have been proposed. Some of the authors have
pointed out that multiple 3-branes could account for the fermion mass
hierarchy.\cite{fermion} In this scenario, the mass hierarchy crucially
depends on the distances between 3-branes. Thus, again, it is
interesting to consider multiple 3-brane configurations with various
distances between 3-branes.
Third, in any higher-dimensional model, stabilizing extra dimensions
is, in general, hard to achieve. In Ref.~\citen{stability} a mechanism for 
stabilizing extra dimensions with multiple branes is proposed.

This paper is organized as follows. In Section 2, we construct static
solutions to Einstein's equations in five spacetime dimensions with any
number of 3-branes put at arbitrary positions. In Section 3, we discuss
the hierarchical structure of our model. In Section 4, we extend the
static solutions found in Section 2 to inflationary
solutions. Conclusions are given in Section 5.

\noindent {\it 2. Many-brane configurations}
\hspace{5mm}
In the Randall-Sundrum model, the orbifold fixed points of $S^1/Z_2$
have been taken as the locations of two 3-branes. Since we would like to
consider any number of 3-branes whose locations are taken to be
arbitrary, we here take a circle $S^1$, rather than $S^1/Z_2$, as the
compactification of an extra dimension.  \footnote{Solutions for
$S^1/Z_2$ may be obtained from those for $S^1$ by imposing the
$Z_2$-symmetry.} The coordinate $\phi$ for the extra dimension is taken
to extend from $0$ to $2\pi$ with the identification of $(x^\mu, \phi=0)$
with $(x^\mu, \phi=2\pi)$.

Let us consider $N$ parallel  3-brane configurations in $5$ spacetime
dimensions. The $i$-th 3-brane may be characterized by the location
$\phi_i$ and the brane tension $V_i$ $(i=1,2,\cdots,N)$. We arrange the
locations of the 3-branes such that $0=\phi_1 < \phi_2 < \cdots < \phi_N
< 2\pi$.  We have here taken the location $\phi_1$ of the first 3-brane
to be the origin of $S^1$ for convenience. Since the $5$-dimensional
spacetime is divided into $N$ domains by $N$ 3-branes, each domain
sandwiched between the $i$-th and the $(i+1)$-th 3-branes can have a
different 5-dimensional cosmological constant $\Lambda_i$.
\footnote{Solutions to the Einstein equations with even numbers of
3-branes are constructed in Ref.~\citen{even}, in which the $\Lambda_i$
have been taken to be identical for all $i$. Solutions similar to ours,
which describe multiple intersecting branes, are given in
Ref.~\citen{junction}}
Thus, the action we start with is given by
\footnote{For the conventions, see the original paper of Randall-Sundrum.
\cite{Randall:1999ee}}
\begin{eqnarray}
S
 &=&S_{{\rm gravity}} + \sum_{i=1}^{N} S_i , \nonumber \\
S_{{\rm gravity}}
 &=&\int d^4x \int_0^{2\pi} d\phi
 \sqrt{-G}
 \left\{
  2 M^3 R - \sum_{i=1}^N \Lambda_i \,
      [ \theta(\phi-\phi_i) - \theta(\phi-\phi_{i+1}) ]
 \right\}, \nonumber \\
S_i
 &=& \int d^4x \sqrt{-g^{(i)}} \left\{ {\cal L}_i - V_i \right\},  
\end{eqnarray}
where $\phi_{N+1} \equiv 2\pi$ and $\theta(\phi)$ denotes the Heaviside
step function defined such that $\theta(\phi)=1$ for $\phi \ge 0$ and
$\theta(\phi)=0$ for $\phi < 0$. 
The quantity $S_i$ is the 4-dimensional $i$-th
3-brane action, and the contribution from the Lagrangian ${\cal L}_i$
will be ignored in the following analysis.

The 5-dimensional Einstein equations for the above action are
\begin{eqnarray}
\sqrt{-G}
\left( R_{MN} - \frac{1}{2}G_{MN} R \right)
&=&
-\frac{1}{4M^3}
\Biggl[
 \sum_{i=1}^N 
  \Lambda_i \, [ \theta(\phi-\phi_i) - \theta(\phi-\phi_{i+1}) ]
  \sqrt{-G}G_{MN}
  \nonumber \\
  && +
 \sum_{i=1}^N V_i \sqrt{-g^{(i)}} g^{(i)}_{\mu\nu} 
 \delta^\mu_M \delta^\nu_N \delta(\phi-\phi_i)
\Biggr].
\end{eqnarray}
Here we solve the equations using the metric
\begin{eqnarray}
 ds^2 = e^{-2\sigma(\phi)}\eta_{\mu\nu} dx^\mu dx^\nu + r_c^2 d\phi^2,
\end{eqnarray}
taken from Ref.~\citen{Randall:1999ee}. Later we attempt
to find solutions describing inflating 3-branes in 5 spacetime
dimensions.

With this form of the metric, the Einstein equations (2) reduce to
\begin{eqnarray}
(\sigma'(\phi))^2 
 &=& -\frac{r_c^2}{24M^3} \sum_{i=1}^N \Lambda_i
     [ \theta(\phi-\phi_i) - \theta(\phi-\phi_{i+1}) ], \\
\sigma''(\phi)
 &=& \frac{r_c}{12M^3} \sum_{i=1}^N V_i \,\delta(\phi-\phi_i). 
\end{eqnarray}
It is not difficult to show that the solution to the above equations has
the form
\begin{eqnarray}
\sigma(\phi)
 &=& (\lambda_1 - 0) (\phi-\phi_1) \theta(\phi-\phi_1) 
    +(\lambda_2 - \lambda_1)(\phi-\phi_2) \theta(\phi-\phi_2)
     \nonumber \\
 & &+ \cdots 
    +(\lambda_N-\lambda_{N-1})(\phi-\phi_N)\theta(\phi-\phi_N),
\end{eqnarray}
where the additive integration constant, which is not physically
relevant, has been chosen for later convenience. 
Since $\sigma(\phi)$ is a function on $S^1$, it has to be periodic,
i.e. $\sigma(2\pi)=\sigma(0)$. This leads to the constraint
\begin{eqnarray}
\sum_{i=1}^N \lambda_i (\phi_{i+1} - \phi_i) = 0.
\end{eqnarray}
The first equation (4) requires
\begin{eqnarray}
\lambda_i = \sqrt{\frac{-\Lambda_i r_c^2}{24M^3}}
\,\,\,\, \mbox{ or } \,\,\,\,
-\sqrt{\frac{-\Lambda_i r_c^2}{24M^3}} .
\,\,\,\,\,\,\,\,\,\,\,\,\,\,\, (i=1,2,\cdots,N)
\end{eqnarray}
We note that every $\Lambda_i$ should be negative for the solution to
make sense, as pointed out in Ref.~\citen{Randall:1999ee}. This
requirement can, however, be relaxed for inflating solutions, as we 
see below. The second equation (5) requires that the 5-dimensional
cosmological constants and the brane tensions be related as
\begin{eqnarray}\
\frac{V_i r_c}{12M^3} = \lambda_i - \lambda_{i-1}, 
\,\,\,\,\,\,\,\,\,\,\,\,\,\,\, (i=1, 2, \cdots, N)
\end{eqnarray}
with $\lambda_0 \equiv \lambda_N$.
\footnote{To obtain the relation for $i=1$, we need to use the
periodicity of $\sigma(\phi)$.}

Although we have given the exact analytical expression for
$\sigma(\phi)$, it may be instructive to determine this geometrically.
The solution (6) turns out to be depicted as follows: First,
specify distinct $N$ points $\phi_i$ which correspond to the locations
of $N$ 3-branes and choose the values of $\sigma(\phi_i)$ at
$\phi=\phi_i$ appropriately. The entire $\phi$-dependence of
$\sigma(\phi)$ can then be obtained by connecting two adjacent points
$\sigma(\phi_i)$ and $\sigma(\phi_{i+1})$ for $i=1,2,\cdots,N$ by
straight lines. In general, each line will be bent at $\phi=\phi_i$. The
slope of the line in the region of $\phi_i \le \phi < \phi_{i+1}$
corresponds to $\lambda_i$, which is related to $\Lambda_i$ through
Eq.~(8). The difference between the slopes of the adjacent lines at
$\phi=\phi_i$ is proportional to the brane tension $V_i$.  Although the
solution $\sigma(\phi)$ can completely be specified by $2N$ parameters
$\phi_i$ and $\sigma(\phi_i)$ ($i=1,2,\cdots,N$), two of them are not
physically relevant. An additive constant to $\sigma(\phi)$ can be
absorbed into an overall constant rescaling of $x^\mu$, and an
overall shift of $\phi_i$ has no physical consequence.

\noindent {\it 3. Hierarchical structure}
\hspace{5mm}
Here we discuss the hierarchical structure of the
solution derived in the previous section. To this end, the mass scale of 
all the parameters ($ \Lambda_i, V_i, M \mbox{ and } r_c $) in the
fundamental theory is assumed to be on the order of the Planck scale.

As mentioned in the previous section, $\sigma(\phi)$ is specified by
$\phi_i$ and $\sigma(\phi_i)$. Without loss of generality, we can assume
that $\sigma(\phi_i) \ge \sigma(\phi_1)=0$. As was done in
Ref.~\citen{Randall:1999ee}, we can derive the 4-dimensional effective
theory by performing the $\phi$ integral. It turns out that the square
of the Planck mass on every 3-brane has the same value, i.e.
\begin{eqnarray}
M_{{\rm Pl}}^2 
 &=& M^3 r_c \int_0^{2\pi} \! \! d\phi \,\,
    e^{-2\sigma(\phi)} \nonumber \\
 &=& M^3 r_c 
  \left[
   \left( \frac{1}{2\lambda_1}-\frac{1}{2\lambda_N} \right)
   +\sum_{i=2}^N 
    \left( \frac{1}{2\lambda_i}- \frac{1}{2\lambda_{i-1}} \right)
    e^{-2\sigma(\phi_i)}
  \right].
\end{eqnarray}
This relation is consistent with the assumption that $M$ and $r_c$ are
on the order of the Planck scale if the $\lambda_i$ are on the order 
of 1 and $\sigma(\phi_i)\gg 1 $ for $i=2,3,\cdots,N$.

Now, the hierarchical structure of our model is obvious. Although the
4-dimensional (effective) Newton constant is observed to be order
$M_{{\rm Pl}}^{-2}$ for every brane, the physical mass scales 
for the $i$-th brane
will reduce by the warp factor $e^{-\sigma(\phi_i)}$ from the
fundamental parameters not far from the Planck scale. The
Randall-Sundrum scenario works well in our model, but we can obtain a
variety of the hierarchy between the Planck scale and physical mass
scales, which is not seen in the original Randall-Sundrum
solution. Thus, in our model the hierarchy problem may be explained as
follows: The reason why the hierarchy between the TeV scale and the
Planck scale is observed in our world is merely that 
we happen to live on a 3-brane whose warp factor is of order
$10^{-15}$.

\noindent {\it 4. Inflating 3-branes} \hspace{5mm} Here, we obtain
solutions describing inflating 3-branes in 5 spacetime dimensions. To
this end, we use the metric \citen{cosmology} and
\citen{Kaloper:1999sm}\footnote{Other parametrizations are also
possible. For example, see Ref.~\citen{another}.}:
\begin{eqnarray}
ds^2 = a(\phi)^2 (-dt^2 + v(t)^2 \delta_{ij} dx^i dx^j) + r_c^2 d\phi^2.
\end{eqnarray}
With this form of the metric, the Einstein equations (2) reduce to 
\begin{eqnarray}
\frac{\ddot{v}(t)}{v(t)}
 &=&
 \left(
   \frac{\dot{v}(t)}{v(t)}
 \right)^2 ,
\\
\left( \frac{a'(\phi)}{a(\phi)} \right)^2
 &=&
 \frac{r_c^2}{a(\phi)^2} 
 \left( \frac{\dot{v}(t)}{v(t)} \right)^2
 - \frac{r_c^2}{24M^3} \sum_{i=1}^N \Lambda_i
 [ \theta(\phi-\phi_i) - \theta(\phi-\phi_{i+1}) ] ,
\\
\frac{a''(\phi)}{a(\phi)}
 &=&
 -\frac{r_c^2}{24M^3}
  \biggl[
   \sum_{i=1}^N \Lambda_i  [\theta(\phi-\phi_i) -\theta(\phi-\phi_{i+1}) ]
    + \frac{2}{r_c} \sum_{i=1}^N V_i \delta(\phi-\phi_i)  
\biggr] ,
\end{eqnarray}
where primes and dots denote derivatives with respect to $\phi$ and $t$, 
respectively. The first equation (12) can easily be solved as
\begin{eqnarray}
v(t) = v(0) e^{Ht},
\end{eqnarray}
where $H$ corresponds to the expansion rate along a 3-brane after
the coefficient of $dt^2$ is normalized to unity on the 3-brane.

Although we have not found a simple expression for $a(\phi)$, like
Eq.~(6), solutions to the above equations turn out to be of the form
\begin{eqnarray}
a(\phi) = \alpha_i e^{\omega_i \phi} + \beta_i e^{-\omega_i \phi}
\,\,\,\,\,\,\,\,\,\,\mbox{ for } \,\, \phi_i \le \phi < \phi_{i+1}. 
\end{eqnarray}
Requiring that $a(\phi)$ be continuous at $\phi=\phi_i$
($i=1,2,\cdots,N$) leads to
\footnote{The continuity condition at $\phi=\phi_1=0$ implies that
$a(2\pi)=a(0)$.}
\begin{eqnarray}
\alpha_N e^{2\pi \omega_N} + \beta_N e^{-2\pi \omega_N}
&=& \alpha_1 + \beta_1, \nonumber
\\
\alpha_{i-1} e^{\omega_{i-1} \phi_i} + \beta_{i-1} e^{-\omega_{i-1} \phi_i}
&=& \alpha_i e^{\omega_i \phi_i} + \beta_i e^{-\omega_i \phi_i}.
\,\,\,\,\,\,\,\,\,\,\,\,\,\,\,\,\,\,\,\,\,(i=2,3,\cdots,N)
\end{eqnarray}
Substituting the expressions (15) and (16) into Eq.~(13) leads to 
\begin{eqnarray}
(\omega_i)^2 &=& - \frac{\Lambda_i r_c^2}{24M^3}, 
\,\,\,\,\,\,\,\,\,\,\,\,\,\,\,\,\,\,\,\,\, (i=1,2,\cdots,N) \\
H^2 &=& \frac{\Lambda_i \alpha_i \beta_i}{6M^3}.
\,\,\,\,\,\,\,\,\,\,\,\,\,\,\,\,\,\,\,\,\, (i=1,2,\cdots,N)
\end{eqnarray}
It may be worth noting that the first relation (18) does not necessarily
imply that $\Lambda_i$ is negative. Solutions may exist even when
$\Lambda_i > 0 $. In this case, $\omega_i$ is purely imaginary, and the
exponential functions in Eq.~(16) should be replaced by trigonometric
functions. Furthermore, even if $H^2$ is negative, we would obtain
physically meaningful solutions by analytic continuation
.\cite{Kaloper:1999sm} The last equation (14) can be satisfied provided
\begin{eqnarray}
&&\left( \omega_i + \frac{V_i r_c}{12M^3} \right)
  \alpha_i e^{\omega_i \phi_i}
-\left( \omega_i - \frac{V_i r_c}{12M^3} \right)
  \beta_i e^{-\omega_i \phi_i}
\nonumber \\
&&=
 \omega_{i-1} \alpha_{i-1} e^{\omega_{i-1} \phi_i}
-\omega_{i-1} \beta_{i-1} e^{-\omega_{i-1} \phi_i},
\,\,\,\,\,\,\,\,\,\,\,\,\,\,\,\,\,\,\,\,\,\,\,(i=1,2,\cdots,N)
\end{eqnarray}
where $\omega_0 \equiv \omega_N$, $\alpha_0 \equiv \alpha_N$ and
$\beta_0 \equiv \beta_N$.

The above solutions for $N=1$ and $2$ (one and two 3-branes) are
investigated in Refs. \citen{cosmology} and \citen{Kaloper:1999sm}. It,
however, seems difficult to analyze the solutions for general $N$
thoroughly. We will not proceed further in this paper. Some of the
physical implications of our solutions may be found in
Refs.~\citen{cosmology} and \citen{Kaloper:1999sm}.

\noindent {\it Conclusion}
\hspace{5mm}
In this paper, we have found new solutions to the Einstein
equations in 5 spacetime dimensions with many 3-branes. The original
Randall-Sundrum solution contains only two 3-branes whose locations are
fixed at the orbifold fixed points of $S^1/Z_2$, and is static. In our
model, any number of parallel 3-branes can be put at arbitrary locations
in the direction of the 5th dimension, and the brane tensions and the
cosmological constants of the 5-dimensional bulks sandwiched between the
3-branes can, in general, be taken to have different values, though they
must satisfy some fine tuning relations for solutions to exist. We
have further succeeded in extending the static solutions to the
inflationary solutions, though our analysis is far from complete.
As in the Randall-Sundrum model, our model can give a solution to the
hierarchy problem between the Planck scale and the TeV scale. Although
the Randall-Sundrum model does not answer other hierarchy problems such
as the fermion mass hierarchy, the existence of multiple 3-branes in our 
model could offer a mechanism to solve them.
A final comment is that although the 5th dimension is compactified on
$S^1$, our solutions will persist on a non-compact space. This can be
seen by simply ignoring the periodicity condition.

\vspace{3mm}

 We would like to thank for C.~S.~Lim for useful comments and discussions.
 The work of M.~T. was supported in part by the Japan Society for the
Promotion of Science.
 K.~T. would like to thank the I.N.F.N. Sezione di Pisa for hospitality.


\end{document}